\documentclass[pra,twocolumn,showpacs]{revtex4}
\usepackage{graphicx}
\usepackage{amsmath}
\usepackage{amssymb}
\usepackage[mathscr]{eucal}
\usepackage{bm}
\usepackage{subfigure}

\allowdisplaybreaks

\newcommand{\beq}{\begin{equation}}
\newcommand{\eeq}{\end{equation}}
\newcommand{\beqa}{\begin{eqnarray}}
\newcommand{\eeqa}{\end{eqnarray}}
\newcommand{\beqan}{\begin{eqnarray*}}
\newcommand{\eeqan}{\end{eqnarray*}}

\newcommand{\tr}[1]{\mathrm{tr} \left( #1 \right) }
\newcommand{\ket}[1]{| #1 \rangle}
\newcommand{\bra}[1]{\langle #1 |}

\newcommand{\ox}{\otimes}

\newcommand{\proof}{\noindent {\bf Proof. }}
\newcommand{\qed}{\hfill $\Box$ \vskip 2ex}

\newtheorem{theorem}{Theorem}

\newtheorem{proposition}{Proposition}

\newtheorem{corollary}{Corollary}

\begin{document}

\title{Reflection Symmetries for Multiqubit Density Operators}
\author{Claudio Altafini}
\affiliation{SISSA-ISAS  \\
International School for Advanced Studies \\
via Beirut 2-4, 34014 Trieste, Italy }

\author{Timothy F. Havel}
\affiliation{Nuclear Science and Engineering \\
MIT, 150 Albany St. \\
Cambridge, MA 02139-4307, USA}



\begin{abstract}
For multiqubit density operators in a suitable tensorial basis, we show that a number of nonunitary operations used in the detection and synthesis of entanglement are classifiable as reflection symmetries, i.e., orientation changing rotations.
While one-qubit reflections correspond to antiunitary symmetries, as is known for example from the partial transposition criterion, reflections on the joint density of two or more qubits are not accounted for by the Wigner Theorem and are well-posed only for sufficiently mixed states.
One example of such nonlocal reflections is the unconditional NOT operation on a multiparty density, i.e., an operation yelding another density and such that the sum of the two is the identity operator. This nonphysical operation is admissible only for sufficiently mixed states.
\end{abstract}

\maketitle 

The Wigner Theorem asserts that unitary and antiunitary operations exhaust all possible symmetric transformations applicable to the wavefunction of a quantum mechanical system.
The unitary transformations are physically associated with forward-in-time evolution, and antiunitary with backward-in-time evolution (see for example Ref.~\cite{Sakurai1}).
The characteristic feature of this last class is the presence of a conjugation operation on a wavefunction or a transposition operation on a density operator.
It is known \cite{Buzek1} that the geometric interpretation of the time reversing operation for a density operator in a two-dimensional Hilbert space (aka ``qubit'') is a \emph{reflection}, i.e.~an orientation-changing rotation in $O^-(3) = O(3) \setminus SO(3)$ of the corresponding Bloch vector.

A closely related operation, variously known as a spin flip \cite{Gisin1}, (unconditional) NOT operation, or universal inverter \cite{Rungta1}, changes the sign of the entire Bloch vector.
In this sense it corresponds geometrically to inversion in the origin, which is widely known as the parity operation \cite{DoranLasenby}.
For a single isolated qubit these operations are indistinguishable from equivalent orientation preserving operations, since $O(3)$ and $SO(3)$ both act transitively on the Bloch sphere, but for multiqubit systems they correspond to partially antiunitary transformations such as the ``partial transposition'', which can be used to detect bipartite entanglement \cite{Peres1}.
This highlights the intrinsically ``discrete'' nature of such tests and their invariance under LOCC (Local Operations and Classical Communication).

In this paper we introduce a more general class of involutory ``symmetry'' operations, and argue that these are likewise useful in studying the multiparty nonseparability of density operators.
These operations are most easily described in terms of the Stokes tensor \cite{Cla-qu-ent1, Cla-spin-tens1, Jaeger1} and its ``unfolding'' to the so-called real density matrix \cite{Havel3}, both of which are equivalent, as carrier spaces, to the coherence vector \cite{Alicki1,MahlerWeberruss}.
All these representations parametrize the real linear space of multiqubit density operators by the expectation values of all possible tensor products of the Pauli operators, differing only in their notations and indexing systems.
The Stokes tensor indexing has the advantage of making the ``affine'' structure of the set of $n$-qubit density operators $\mathcal D_n$ explicit, whereas the real density matrix has the advantages that both the matrix itself, as well as any operations on it which are diagonal with respect to the Stokes tensor, can be displayed as a compact 2-D table on a printed page (see below for examples).

As is well-known, unitary operations on the usual Hermitian density operator induces orientation-preserving rotations of the coherence vector, and thereby also norm-preserving linear group actions on the Stokes tensor / real density matrix [\emph{loc.~cit.}].
In the following, we shall frequently use the term ``density'', without further qualification, to indicate an equivalence class of probability distributions over an ensemble of multiqubit systems which all give rise to the same density operator, irrespective of how this is represented (as a Hermitian matrix, or a Stokes tensor, etc.).

We now distinguish the following two types of nonunitary but norm-preserving operations on a multiqubit Stokes tensor:
\begin{description}
\item[(i)] local reflections applied simultaneously to two or more qubits; 
\item[(ii)] ``nonlocal'' reflections, i.e.\ reflections applied to the \emph{joint} density of two or more qubits.
\end{description}
The two cases are \emph{qualitatively different}: while (i) is equivalent, up to local unitary operations, to multiqubit partial transposition, (ii) is a genuinely new operation and does not correspond to any local operation on two or more qubits.
In particular, the total reflection of all components of the Stokes tensor other than the expectation value of the identity does \emph{not} correspond to time-reversal (i.e.\ to the total transpose of the density matrix) but rather to a multiparty NOT operation.

Reflections on more than one qubit are nonunitary operations that do not necessarily yield valid (positive semidefinite) density operators.
However, it can be shown that any mixed state with eigenvalues ``small enough''
is still a density operator when it is totally reflected.
In other words, total reflection is a nonunitary involution which preserves such sufficiently mixed sets of density operators.
On this set, total reflection behaves like a anti-unitary operation in the sense that it preserves the Hermitian structure, the trace and the (Hilbert-Schmidt) inner product.
This tells us that for general mixed states there are more symmetries to be exploited than those of Wigner theorem.

For three qubits, the set of density operators admitting a total reflection includes for example the Unextendible Product Basis (UPB) states used in Ref.~\cite{Bennett1} to generate a bound entangled density operator with all positive partial transpositions (PPT).
The ``complement'' operation that turns a separable density into the bound entangled UPB state is in fact a total reflection of the type (ii) above.
The various entanglement measures (the concurrence, the negativity and the tangle among them) that rely on the use of spin-flip operations are also examples of application of multiple one-qubit reflections of the type~(i).
In between local and total reflections lies a class of ``nonlocal yet partial'' reflections which also belong to the class (ii) above. 
These maps resemble very closely those used in the so-called reduction criterion \cite{Cerf1,Horodecki8}.

Besides their unifying mathematical (group theoretic) character, we see reflections as a new tool to ``probe'' the structure of the set of multiparticle density operators, in particular its nonseparable regions, by means of operations analogous, but inequivalent, to partial transposition. Hopefully this will eventually lead to a better understanding of bound entanglement in multipartite systems.

\section{One qubit: Transposition and time-reversal}
For a single qubit with density operator $\rho \in \mathcal D_1 \subset \mathbb C^{2\times2}$, the Stokes tensor is the affine 3-vector $[\varrho^0 \, \smash{\vec{\varrho}\,}^T]^T$, where  $\varrho^0 = \tr{\rho}/\sqrt2 = 1/\sqrt{2}$ and $\vec{\varrho} =  [ \varrho^1\, \varrho^2 \,\varrho^3 ]^T$ is the Bloch vector of the qubit times $\sqrt2$. Thus (summing over repeated indices) we have $\rho = \varrho^j \lambda_j$ where $\lambda_j =  \sigma_j /\sqrt{2}$ are the rescaled Pauli matrices ($j=1, \, 2 , \, 3$), $\lambda_0 =  \openone_2 / \sqrt2$, and $\varrho^j = \tr{\rho \lambda_j }$.
In this notation, the real density matrix is given by
\begin{equation}
\sigma = \sigma(\rho) ~\equiv~ \sqrt2 \begin{bmatrix} \varrho^0 & \varrho^2 \\[0.75ex] \varrho^1 & \varrho^3 \end{bmatrix} .
\end{equation}
The Stokes tensor is readily recovered from this by applying the ``\textrm{col}'' \cite{Havel2} (aka reshaping \cite{ZyczkowskiBengtsson}) operator to it and dividing by $\sqrt2$.
As is well-known, unitary transformations of $\rho$ by $U \in SU(2)$, namely $U\rho\,U^\dagger$, induce rotations of the corresponding Bloch vector.
This geometric interpretation will now be extended to antiunitary transformations \cite{Busch1,Lewenstein1}.

Any antiunitary operation can be written as the product of a unitary operation and complex conjugation $K$. 
Given a pure state with wave vector $\ket{\psi} =  c_0 \ket{0} + c_1 \ket{1}$ ($c_0, c_1 \in \mathbb{C}$), let $\ket{\tilde{\psi}}$ be the wave vector obtained by means of $K$ alone: $\ket{\tilde{\psi}} = K \ket{\psi}  =  c_0^\ast \ket{0} + c_1^\ast \ket{1}$. 
The corresponding density matrix is $\rho = \ket{\psi} \bra{\psi} = \sum_{j,k=0}^{1,1} c_j c_k^\ast \ket{j} \bra{k}$, so that $\ket{\tilde{\psi}} \bra{\tilde{\psi}} = \sum_{j,k=0}^{1,1} c_j^\ast c_k \ket{j} \bra{k} = ( \ket{\psi} \bra{\psi} )^T$.
Since any density matrix is a convex combination of pure state density matrices, the effect of $K$ on a general $\rho$ is to transpose it, i.e.~$\rho^T = K \rho K^\dagger = \varrho^0 \lambda_0 + \varrho^1 \lambda_1 - \varrho^2 \lambda_2 + \varrho^3 \lambda_3$.  As indicated, this is simply a change in the sign of the $\lambda_2$ component of the Bloch vector, i.e.~$[\varrho^1 \,  -\! \varrho^2 ~ \varrho^3 ]^T$.

\begin{figure}[ht]
\begin{center}
 \includegraphics[width=3.5cm]{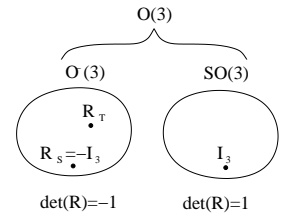}
 \caption{Topological structure of the rotation group $O(3)$.}
\label{fig:O3-SO3}
\end{center}
\end{figure}

The rotation group $O(3)$, of course, has two connected components, one of which preserves the orientation of a frame (namely $SO(3)$, which contains the identity operator $\openone_3$), and one of which changes its orientation (denoted here by $O^-(3)$, to which $ - \openone_3$ belongs).
This topological structure is illustrated in Fig.~\ref{fig:O3-SO3}.
A \emph{reflection} is a rotation which does not preserve orientation.
The canonical example is \emph{spatial inversion}, which is defined as multiplication by $R_S \equiv -\openone_3$.
Any reflection $R \in O^-(3)$ is obtained by multiplying $R_S$ with a rotation in $SO(3)$.
For example, the reflection used in the transpose, $R_T  = \mathrm{diag}(  1 ,\,  -1 ,\,  1 )$, can be written as the product of a spatial inversion with a rotation by $\pi$ about the $y$-axis.

For any vector $\vec{\varrho}$, spatial inversion maps $\vec{\varrho}$ to its antipode $-\vec{\varrho} = R_S (\,\vec{\varrho}\,)$ on a sphere of radius $\| \vec{\varrho}\, \| = \sqrt{(\varrho^1)^2 +(\varrho^2)^2 +(\varrho^3)^2 }$.
It follows from this together with the above that, for density matrices, $\rho^S \equiv U K \rho K^\dagger U^\dagger =\varrho^0 \lambda_0 - \varrho^1 \lambda_1 - \varrho^2 \lambda_2 - \varrho^3 \lambda_3$ where $U = \imath \lambda_2 \in SU(2)$ rotates the Bloch vector by $\pi$ about the $y$-axis.
In addition, it is easily shown that the eigenvalues of $\rho$ are given by
\beq
\mathrm{eig}\left( \rho \right) = \left\{ \tfrac{1}{\sqrt{2}}\! \left( \tfrac{1}{\sqrt{2}} \pm \| \vec{\varrho}\, \| \right)   \right\} .
\label{eq:eig:rho}
\eeq
Since reflections, like rotations in $SO(3)$, are length preserving actions on the Bloch sphere, we see that the eigenvalues are preserved under reflections: $\mathrm{eig}(\rho^S) = \mathrm{eig}(\rho^T) = \mathrm{eig}(\rho)$.
For pure states, an important difference between $R_S$ and $R_T$ is that $R_S$ maps any ket $\ket{\psi}$ to an orthogonal one, whereas $R_T$ does not.
In other words, spatial inversion corresponds exactly to the spin-flip operation \cite{Gisin1, Jaeger2}.

Both the transposition and the spin-flip can also be defined in terms of the real density matrix, using the component-wise (aka Hadamard, or sometimes Schur) matrix product ``$\odot$''.
In the case of the transpose, this is simply:
\beq
\sigma( \rho^T ) ~=~ \sqrt2 \begin{bmatrix} \;1 & -1 \\[0.75ex] \;1 & 1 \end{bmatrix} \odot \begin{bmatrix} \varrho^0 & \varrho^2 \\[0.75ex] \varrho^1 & \varrho^3 \end{bmatrix} .
\eeq
As shown in Ref.~\cite{Havel2}, an operator sum representation is obtained from the singular value decomposition of the sign matrix (left-hand factor), leading to
\beq
\sigma( \rho^T ) ~=~ \sigma( \rho ) \ket0\bra0 - \sqrt2\, \lambda_3\, \sigma( \rho ) \ket1\bra1 ~.
\eeq
For the spin flip, on the other hand, it is easily seen that
\beq
\sigma( \rho^S ) \begin{aligned}[t] ~=~ &
\sqrt2 \begin{bmatrix} 1 & -1 \\[0.75ex] -1 & -1 \end{bmatrix} \odot \begin{bmatrix} \varrho^0 & \varrho^2 \\[0.75ex] \varrho^1 & \varrho^3 \end{bmatrix} \\[1ex]
 ~=~ & 2 \ket{0}\bra{0} - \sigma( \rho ) \end{aligned}
\eeq
These alternative representations of transposition and spin flip will be useful in studying multiqubit reflections below.

For a single qubit the notion of reflection admits a further interpretation in terms of ``co-completely positive'' (co-CP) maps.
From the St{\o}rmer-Woronowicz theorem, any positive $ 2\times 2$ map $ \Phi $ is decomposable as 
\beq
\Phi \;=\; c\, \Phi_1 + (1 - c) \Phi_2 \circ T \qquad (0 \leqslant c \leqslant 1) ,
\label{eq:deco}
\eeq
where $ \Phi_1$, $\Phi_2$ are completely positive (CP) maps and $T$ is transposition.
The composition $\Phi_2 \circ T $ is called a \emph{co-CP map}.
For the Bloch vector, the CP maps form a semigroup in the group of orientation-preserving affine maps $GL^+ (3, \, \mathbb{R} ) \,\circledS\, \mathbb{R}^3 $, where $  GL^+ (3, \, \mathbb{R} ) = \{ g \in GL(3, \, \mathbb{R}^3 ) \mid \det (g) > 0 \} $ and ``$\circledS$'' denotes its semi-direct product with the translation group $\mathbb{R}^3$ \cite{ZyczkowskiBengtsson, Ruskai1, Cla-contr-open1}.
Unital CP maps live in the $GL^+(3, \, \mathbb{R})$ component, while unital co-CP maps live in the other component, $GL^- (3, \, \mathbb{R} ) \equiv \{ g \in GL(3, \, \mathbb{R}^3 ) \mid \det (g) < 0 \} $.
Restricting further to symmetries (i.e.~trace- and norm-preserving maps), one gets rotations and reflections as above.

\section{Two qubits: partial transposition, partial time reversal, multiple local reflections and total reflections}
For two qubits, a complete basis for the space of density matrices $\mathcal D_2 \subset \mathbb{C}^{4\times4}$ is given by $\Lambda_{jk} = \lambda_j \otimes \lambda_k$  ($j, \,k \in \{ 0, \, 1 ,\, 2,\, 3\}$).
This basis is also orthonormal relative to the Hilbert-Schmidt inner product, i.e.~$\tr{\Lambda_{jk}\Lambda_{lm} }=  \delta_{jl}\delta_{km}$ for all $j, \, k, \, l, \, m \in \{ 0, \, 1, \, 2, \, 3 \}$.
For a given density matrix $\rho$, the $\Lambda$-basis defines a real, rank 2 tensor $ \varrho^{jk}$ which gives a contravariant representation of the same density: $ \rho = \varrho^{jk} \Lambda_{jk}$.
Viewed as a 16-vector, $\varrho^{jk} $ is affine, i.e.~$\varrho^{00} = \tr{\rho \Lambda_{00} } = 1/2$, and
it is bounded by the 15-dimensional sphere in $ \mathbb{R}^{16}$ of radius $1$,

\beq
\tr{\rho^2}  = \tr{ \left( \varrho^{jk} \Lambda_{jk}\right)^2 }
= \sum_{j, k=0}^{3,3} \left( \varrho^{jk} \right)^2 \le 1 ,
\label{eq:Casim1}
\eeq
with equality if and only if the state is pure.

A two-qubit density matrix $\rho$ is said to be \emph{separable} if it can be written as a convex combination $\rho = \sum_{r=1}^s  w^r \rho_{A, r} \otimes \rho_{B,r}$ for some set of real numbers $w^r \geqslant 0$ such that $\sum_{r=0}^s  w^r = 1$, where $\rho_{A, r}$, $\rho_{B,r}$ are all single-qubit density matrices.
A necessary and sufficient condition for the separability of a two-qubit density is provided by the positive partial transpose (PPT) criterion of Peres \cite{Peres1} and Horodecki \cite{Horodecki1}.
The partial transpose of a two-qubit density matrix $\rho$ with respect to the first (left) subsystem $A$ is defined as $\rho^{T_A}  \equiv \left( K \otimes \openone_2 \right) \rho \left( K \otimes \openone_2 \right)^\dagger$, and similarly $\rho^{T_B} \equiv \left( \openone_2 \otimes K \right) \rho \left(\openone_2 \otimes K \right)^\dagger$.
Each partial transpose is still a well-defined (i.e.~positive semidefinite) density operator if and only if $ \rho$ is separable.
The PPT criterion may be viewed as check on the feasibility of the ``partial time reverse'' operation \cite{Busch1,Lewenstein1}: changing the time arrow of one of the subsystems alone.

In terms of the Stokes tensor $\varrho^{\,jk} $, the description of partial transposition is very intuitive and relies on the observation that $\lambda_2 = -\lambda_2^T$ is the only Pauli matrix with imaginary elements.
\begin{proposition}
\label{prop:PPT2}
For two qubits, the partial transpose operations on the density matrix $\rho^{T_A}$ and $\rho^{T_B}$
act on the Stokes tensor $\varrho^{jk}$ by changing the sign of all elements bearing the index ``2'' in the corresponding subsystem:
\begin{subequations}
\label{eq:PPTvectch}
\beqa
\rho^{T_A} & = & \varrho^{0k} \Lambda_{0k} +  \varrho^{1k} \Lambda_{1k} -  \varrho^{2k} \Lambda_{2k} +  \varrho^{3k} \Lambda_{3k} 
\label{eq:partT1vectch}\\
\rho^{T_B} & = & \varrho^{\,j0} \Lambda_{\,j0} +  \varrho^{\,j1} \Lambda_{j1} -  \varrho^{\,j2} \Lambda_{j2} +  \varrho^{\,j3} \Lambda_{j3}
\label{eq:partT2vectch}
\eeqa
\end{subequations}
\end{proposition}
The verification is just a straightforward calculation, which may be found in Table~\ref{tab:refl-2qubit} below.
Note also that for the ``total'' transpose $ \rho^T$ ($=(\rho^{T_A} )^{T_B}$) we have instead
\begin{subequations}
\label{eq:PPTvectch-proof}
\beqa
\Lambda_{jk}^T & =  &\Lambda_{jk}  \qquad \text{if $j, k \neq 2$ or $j=k=2$}
\label{eq:partT1vectch-proof} \\
\Lambda_{jk}^T & = & - \Lambda_{jk}  ~\quad \text{if $j=2$ or $k = 2$, $ j\neq k$} ,
\label{eq:partT2vectch-proof}
\eeqa
\end{subequations}
showing that $\Lambda_{22}$ behaves differently under partial or total transposition.

The PPT separability test of Peres-Horodecki relies essentially on the decomposability property \eqref{eq:deco}: any 1-qubit positive but not CP map, when applied to a 2-qubit density, returns a density if and only if the original density is separable.
Restricting from positive maps to symmetry operations is the same as restricting to local reflections.
In fact, the map \eqref{eq:partT1vectch} can be thought of as the linear transformation $\bar{R}_T \otimes \openone_{4} $, where $ \bar{R}_T $ is the following affine orientation-changing three-dimensional rotation: $ \bar{R}_T = \mathrm{diag} \left(  1, \,  R_T \right) =\mathrm{diag} \left(  1, \, 1, \, -1, 1 \right) $.
Since all single qubit reflections are unitarily equivalent, any matrix $ R \in O^-(3)  $ can be used in place of $R_T$.
Indeed, if $\bar{R} = \mathrm{diag} \left(  1, \,  R \right)$, then local operations from the same connected component of $O(3)$ satisfy
\beq
\mathrm{eig} \left( \left(  \bar{R} \otimes \openone_{4} \right)   \left( \rho \right) \right) = \mathrm{eig} \left( \left( \bar{R}_T \otimes \openone_{4} \right) ( \rho )\right)  ,
\label{eq:eigR}
\eeq
where the notation must be interpreted as follows: the matrix $ \bar{R} \otimes \openone_{4} $ acts on the 16-vector $\varrho^{jk}$ and the resulting 16-vector provides the coefficients in the sum over the basis elements $\Lambda_{jk}$, i.e. $( \bar{R} \otimes \openone_{4}) ( \rho ) = \left( \bar{R} \otimes \openone_{4})^{lm}_{jk}   \varrho^{jk} \right) \Lambda_{lm}$.

Eq.~\eqref{eq:eigR} shows that all reflections are positive but not completely positive.
Thus we can reformulate the PPT criterion for the separability of two qubits as follows:
\begin{theorem}
A two-qubit density matrix $\rho$ is separable if and only if $\left( \bar{R} \otimes \openone_{4} \right) ( \rho) $ is a density matrix for any $R \in O^-(3)$.
\end{theorem}
A particularly simple such map is $\bar{R}_S = \mathrm{diag} ( 1, \,R_S )$, where $R_S$ is the spin flip operation from the previous section. It is easily seen that $\left(  \bar{R}_S \otimes \openone_{4}  \right) (\rho) = 2 \varrho^{0k} \Lambda_{0k} - \rho$, so that the sign is changed in all elements $\varrho^{jk}$ except those appearing in the reduced density matrix of the second qubit (i.e.~the $\varrho^{0k}$).

\begin{figure}[ht]
\begin{center}
\subfigure[ ]{
 \includegraphics[width=6cm]{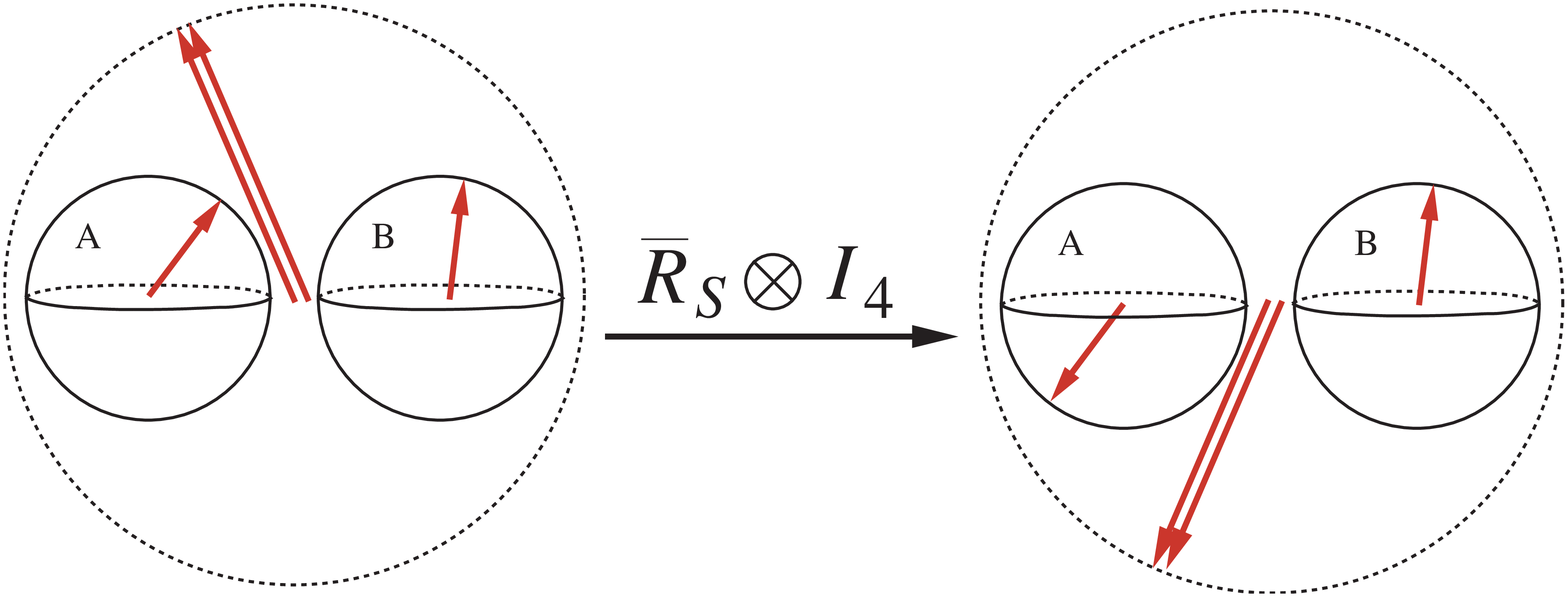}}
\subfigure[ ]{
 \includegraphics[width=6cm]{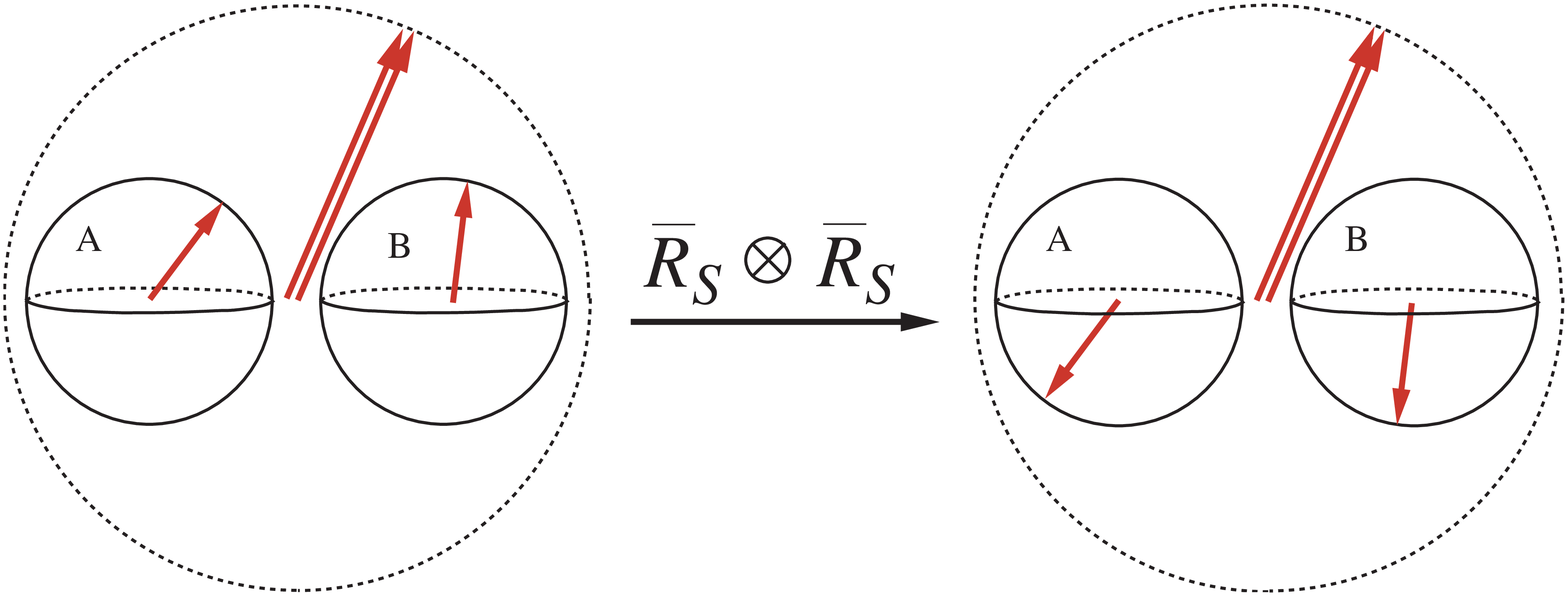}}
\subfigure[ ]{
 \includegraphics[width=6cm]{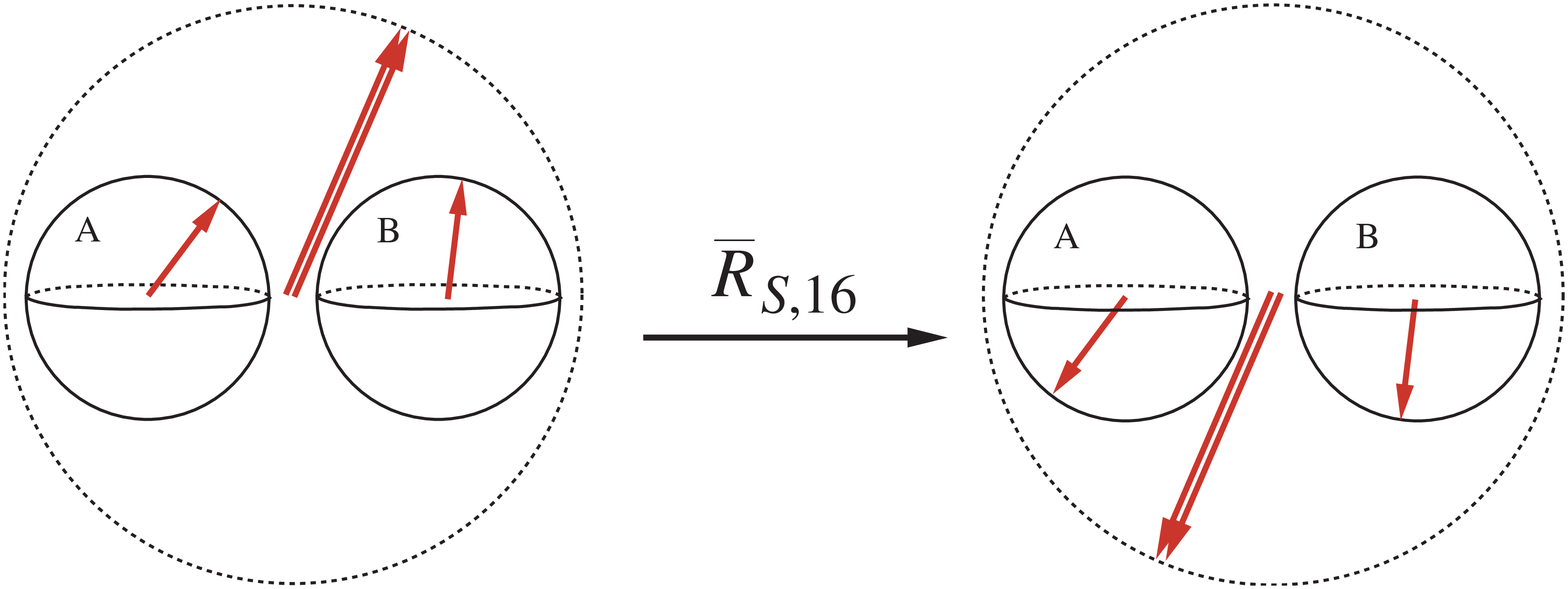}}
 \caption{Reflections on a 2-qubit density matrix. The two vectors contained in the smaller spheres correspond to the Bloch vectors $ \varrho^{j0} $ and $ \varrho^{0k} $ of the two reduced density matrices, the third vector (double arrow) to the 2-body correlation part of the Stokes tensor $\varrho^{jk}$, $j,k\neq 0$: (a) the single qubit reflection $\bar{R}_S\ox \openone_4$ (PPT test); (b) the double local reflection $\bar{R}_S \ox \bar{R}_S$ (which is equivalent to the total transpose under LOCC); (c) the total reflection $\bar{R}_{S,16}$ (a nonlocal operation).}
\label{fig:Stokes-refl}
\end{center}
\end{figure}

The (total) transpose $\rho^T$ of $\rho$ corresponds to the matrix $\bar{R}_{T,16}= \bar{R}_T\ox \bar{R}_T = \mathrm{diag} ( 1, \, R_{T, 15} )$ with $ R_{T, 15} = \mathrm{diag} ( 1, -1, ~1, ~1,~1,-1,~1, -1, -1,~1, -1,~1,~1, -1,~1) \in SO(15)$, \linebreak[2] where the minus signs correspond to the 6 basis elements obeying \eqref{eq:partT2vectch-proof}.
Since the determinant of this matrix is positive, for two qubits the transpose is an orientation-preserving operation.
Up to local operations $\bar{R}_T \ox \bar{R}_T$ is equivalent to the ``double local reflection'' (or double spin flip) map $\bar{R}_S \ox  \bar{R}_S$.
The difference between $ \bar{R}_S  \ox \openone_4 $ and $ \bar{R}_S \ox \bar{R}_S$ is easily understood by looking at Fig.~\ref{fig:Stokes-refl}.
While $\bar{R}_S \otimes \openone_4$ leaves the reduced density of the second qubit unchanged (Fig.~\ref{fig:Stokes-refl}(a)), the correlation part remains unchanged under the action of $\bar{R}_S \ox \bar{R}_S$ because its sign is flipped twice (Fig.~\ref{fig:Stokes-refl}(b)).
It may be shown, however, that both are positive but not-completely-positive maps.

All the ``local'' maps in $O(3)$ mentioned so far are orientation-preserving when acting on two qubits, even though they all have at least one factor that is orientation-changing when acting on a single, isolated qubit: $ 
 \det(\bar{R}_T \ox  \openone_4) = \det(\bar{R}_T \ox \bar{R}_T ) = \det(\bar{R}_S  \ox \openone_4 ) = \det(\bar{R}_S \ox  \bar{R}_S) =1 $. 
The recovery of ``parity'' whenever an orientation-changing map is applied to two or more qubits is due to the affine structure of the Hilbert space of a qubit (resulting in a \emph{affine Bloch vector}), itself a consequence of the trace-preserving condition:
\[
\begin{bmatrix} 1 & \\ & O(3) \end{bmatrix} \ox \begin{bmatrix} 1 & \\ & O(3) \end{bmatrix} ~\subset~ \begin{bmatrix} 1 & \\ & SO(15) \end{bmatrix}, 
\]
Hence the question arises: do there exist any orientation-changing symmetric operations on two qubits?
One such map is the 2-qubit total reflection $\bar{R}_{S,16} = \mathrm{diag}( 1, -\openone_{15} )$, $-\openone _{15} \in O(15)\setminus SO(15) = O^-(15)$. 
Its action (see Fig.~\ref{fig:Stokes-refl}(c)) corresponds to changing the sign of the entire tensor $\varrho^{jk}$ except for affine component $\varrho^{00} = 1/2$, thus the name \emph{total reflection}.
This nonlocal operation is genuinely new and inequivalent to any composition of local symmetric operations.

The most significant difference between total transpose and total reflection is that whereas the former map preserves the eigenvalues of the density matrix, the latter does not.
Indeed, the total reflection is not even a positive map, since it converts the density matrix of any pure state to one with eigenvalues $[1,\, 1,\, 1,\, -1] / 2$.
This fact is readily established by writing the total reflection directly in terms of the Hermitan density matrix as
\beq
\bar{R}_{S,16}( \rho ) ~=~ \tfrac12\, \openone_4 \,-\, \rho ~,
\eeq
which makes it clear that it holds for the density matrices of the basis states $\ket{00}\bra{00}, \ket{01}\bra{01}, \ket{10}\bra{10}$ \& $\ket{11}\bra{11}$, and that the total reflection commutes with arbitrary two-sided unitary transformations of $\rho$.

The changes in the signs of the elements of the Stokes tensor are summarized in Table~\ref{tab:refl-2qubit} for all the discrete symmetric operations mentioned in this Section.
It may be observed that $\bar{R}_S \ox  \bar{R}_S$ and $\bar{R}_{T,16} = \bar{R}_T \otimes \bar{R}_T$ both have an even number of ``$-$'' signs (6), whereas $\bar{R}_{S,16}$ has an odd number (namely 15), thus confirming that a total reflection on a two-qubit joint density is inequivalent to such operations.

\begin{widetext}

\begin{table}[ht!]
\caption{Action (sign changes) of the rotations and reflections involving  $\bar{R}_{T}$, $\bar{R}_S$ and $\bar{R}_{S,16}$ on the components of the 2-qubit Stokes tensor $\varrho^{jk}$.}
\begin{center}
\begin{tabular}{|c|c|c|c|c|c|c|c|}
\hline
 $ \varrho^{jk} $  &  $\bar{R}_{T}\ox   \openone_4 $ &  $\openone_4 \ox \bar{R}_{T}$ &  $\bar{R}_{T} \ox \bar{R}_{T} $ & $\bar{R}_S \ox \openone_4$ & $\openone_4 \ox \bar{R}_S$ &  $\bar{R}_S \ox \bar{R}_S$  &  $\quad \bar{R}_{S,16} \quad$   \\
\hline
\hline
$ \varrho^{00} $ & $ + $ & $ + $ & $ + $ & $ + $ & $ + $ & $ + $ & $ + $ \\
 \hline
$ \varrho^{01} $ & $ + $ & $ + $ & $ + $ & $ + $ & $ - $ & $ - $ & $ - $ \\
\hline
$ \varrho^{02} $ & $ + $ & $ - $ & $ - $ & $ + $ & $ - $ & $ - $ & $ - $ \\
 \hline
$ \varrho^{03} $ & $ + $ & $ + $ & $ + $ & $ + $ & $ - $ & $ - $ & $ - $ \\
\hline
$ \varrho^{10} $ & $ + $ & $ + $ & $ + $ & $ - $ & $ + $ & $ - $ & $ - $ \\
 \hline
$ \varrho^{11} $ & $ + $ & $ + $ & $ + $ & $ - $ & $ - $ & $ + $ & $ - $ \\
\hline
$ \varrho^{12} $ & $ + $ & $ - $ & $ - $ & $ - $ & $ - $ & $ + $ & $ - $ \\
 \hline
$ \varrho^{13} $ & $ + $ & $ + $ & $ + $ & $ - $ & $ - $ & $ + $ & $ - $ \\
 \hline
$ \varrho^{20} $ & $ - $ & $ + $ & $ - $ & $ - $ & $ + $ & $ - $ & $ - $ \\
\hline
$ \varrho^{21} $ & $ - $ & $ + $ & $ - $ & $ - $ & $ - $ & $ + $ & $ - $ \\
 \hline
$ \varrho^{22} $ & $ - $ & $ - $ & $ + $ & $ - $ & $ - $ & $ + $ & $ - $ \\
\hline
$ \varrho^{23} $ & $ - $ & $ + $ & $ - $ & $ - $ & $ - $ & $ + $ & $ - $ \\
 \hline
$ \varrho^{30} $ & $ + $ & $ + $ & $ + $ & $ - $ & $ + $ & $ - $ & $ - $ \\
 \hline
$ \varrho^{31} $ & $ + $ & $ + $ & $ + $ & $ - $ & $ - $ & $ + $ & $ - $ \\
 \hline
$ \varrho^{32} $ & $ + $ & $ - $ & $ - $ & $ - $ & $ - $ & $ + $ & $ - $ \\
 \hline
$ \varrho^{33} $ & $ + $ & $ + $ & $ + $ & $ - $ & $ - $ & $ + $ & $ - $ \\
\hline
\hline
\# sign changes & $ 4 $ & $ 4 $ & $ 6 $ & $ 12 $ & $ 12 $ & $ 6 $ & $ 15 $ \\
\hline
\end{tabular}
\label{tab:refl-2qubit}
\end{center}
\end{table}

\end{widetext}

In terms of density matrices, the positive-but-not-completely-positive operation $\bar{R}_S \ox  \bar{R}_S$ corresponds to 
\beq
\bar{R}_S \ox  \bar{R}_S(\rho) = (\sigma_2 \ox \sigma_2) \rho^\ast (\sigma_2 \ox \sigma_2) = 4 \Lambda_{22}\, \rho^\ast \Lambda_{22} .
\eeq
The transformed density matrix $\rho' \equiv \bar{R}_S \ox  \bar{R}_S (\rho)$ is frequently found in entanglement measures, including the concurrence $C(\rho) = \max \{ 0, \nu_1-\nu_2-\nu_3-\nu_4 \}$ (where $\nu_j \in \mathrm{eig} (\rho\rho' )$ \cite{Wootters2}) and the Lorentzian metric $ \tr{ \rho \rho' } = (\varrho^{00} ) ^2 - \sum_{j=1}^3 \left( (\varrho^{0j} )^2 + (\varrho^{j0} )^2 \right) + \sum_{j,k=1}^3 ( \varrho^{jk} )^2$ \cite{Jaeger1}.

\section{Two Qubits: Matrix Structures and the Computable Cross-Norm}

In this section, we show how the foregoing nonunitary symmetry operations on a two-qubit density matrix can be expressed compactly using the Hadamard product of matrices \cite{Horn1} together with either the Stokes tensor or the real density matrix.
We will also show that a non-separability criterion called the computable cross-norm \cite{Rudolph} (or the matrix realignement method \cite{Chen2}), which is inequivalent to the PPT criterion, can be computed directly from the Stokes tensor without having to convert back to the traditional Hermitian representation.
For two qubits, the Stokes tensor can also be viewed as a square array of real numbers,  which is related to the real density matrix as follows:
\beq
2\! \begin{bmatrix}
\varrho^{00} & \varrho^{01} & \varrho^{02} & \varrho^{03} \\[1ex]
\varrho^{10} & \varrho^{11} & \varrho^{12} & \varrho^{13} \\[1ex]
\varrho^{20} & \varrho^{21} & \varrho^{22} & \varrho^{23} \\[1ex]
\varrho^{30} & \varrho^{31} & \varrho^{32} & \varrho^{33}
\end{bmatrix}
\longleftrightarrow
\begin{bmatrix}
\varrho^{00} & \varrho^{20} & \varrho^{02} & \varrho^{22} \\[1ex]
\varrho^{10} & \varrho^{30} & \varrho^{12} & \varrho^{32} \\[1ex]
\varrho^{01} & \varrho^{21} & \varrho^{03} & \varrho^{23} \\[1ex]
\varrho^{11} & \varrho^{31} & \varrho^{13} & \varrho^{33}
\end{bmatrix} .
\label{eq:tensor2matrix}
\eeq
The rearrangement of the elements seen here corresponds to the Choi \cite{Havel2} (aka reshuffling \cite{ZyczkowskiBengtsson}) map for $n = 2$ qubits, but for $n > 2$ the Stokes tensor-to-real density matrix map is not the same as the Choi map; indeed, then the order of the Stokes tensor is greater than two, so it can no longer be identified so simply with a matrix.

The real density matrix has the useful feature of preserving the tensor product structure of the corresponding Hermitian density matrix, i.e.~for two qubits: $\sigma( \rho \ox \rho') = \sigma(\rho) \ox \sigma(\rho') \equiv \sigma \ox \sigma'$.
It follows immediately that a 2-qubit real density matrix can be written as a convex combination of 1-qubit real density matrices if and only if the 2-qubit density is separable.
A $2\times2$ real matrix, on the other hand, is a real density matrix if and only if its upper-left element is unity and the length of the Bloch vector determined by the remaining elements does not exceed unity (cf.\ Eq.\ \ref{eq:eig:rho}).
It should also be noted that, with either the real density matrix or the Stokes tensor, the partial trace operation involves only discarding elements involving the qubit traced over: no additional operations are needed as in the Hermitian representation.

As shown previously for the 1-qubit case, we can express involutory symmetry operations by means of Hadamard products of the real density matrix with matrices the elements of which are all $\pm1$.
Moreover, these matrices will be tensor products if and only if the operations that define them are.
This may be seen in the following list of sign matrices for all the operations given in Table \ref{tab:refl-2qubit}:
\begin{subequations} \beqa
\bar{R}_{T} \ox \openone_4 &\leftrightarrow&
\left[ \begin{smallmatrix} &\\[0.5ex] +1 & -1 \\[1ex] +1 & +1\\[-0.25ex] & \end{smallmatrix} \right] \otimes
\left[ \begin{smallmatrix} &\\[0.5ex] +1 & +1 \\[1ex] +1 & +1\\[-0.25ex] & \end{smallmatrix} \right] \\
\openone_4 \ox \bar{R}_{T} &\leftrightarrow&
\left[ \begin{smallmatrix} &\\[0.5ex] +1 & +1 \\[1ex] +1 & +1\\[-0.25ex] & \end{smallmatrix} \right] \otimes
\left[ \begin{smallmatrix} &\\[0.5ex] +1 & -1 \\[1ex] +1 & +1\\[-0.25ex] & \end{smallmatrix} \right] \\
\bar{R}_{T} \ox \bar{R}_{T} &\leftrightarrow&
\left[ \begin{smallmatrix} &\\[0.5ex] +1 & -1 \\[1ex] +1 & +1\\[-0.25ex] & \end{smallmatrix} \right] \otimes
\left[ \begin{smallmatrix} &\\[0.5ex] +1 & -1 \\[1ex] +1 & +1\\[-0.25ex] & \end{smallmatrix} \right] \\
\bar{R}_S \ox \openone_4 &\leftrightarrow&
\left[ \begin{smallmatrix} &\\[0.5ex] +1 & -1 \\[1ex] -1 & -1\\[-0.25ex] & \end{smallmatrix} \right] \otimes
\left[ \begin{smallmatrix} &\\[0.5ex] +1 & +1 \\[1ex] +1 & +1\\[-0.25ex] & \end{smallmatrix} \right] \\
\openone_4 \ox \bar{R}_S &\leftrightarrow&
\left[ \begin{smallmatrix} &\\[0.5ex] +1 & +1 \\[1ex] +1 & +1\\[-0.25ex] & \end{smallmatrix} \right] \otimes
\left[ \begin{smallmatrix} &\\[0.5ex] +1 & -1 \\[1ex] -1 & -1\\[-0.25ex] & \end{smallmatrix} \right] \\
\bar{R}_S\ox \bar{R}_S &\leftrightarrow&
\left[ \begin{smallmatrix} &\\[0.5ex] +1 & -1 \\[1ex] -1 & -1\\[-0.25ex] & \end{smallmatrix} \right] \otimes
\left[ \begin{smallmatrix} &\\[0.5ex] +1 & -1 \\[1ex] -1 & -1\\[-0.25ex] & \end{smallmatrix} \right] \\
\bar{R}_{S,16} &\leftrightarrow&
\left[ \begin{smallmatrix} &&& \\[0.5ex]  +1 & -1 & -1 & -1 \\[1ex]  -1 & -1 & -1 & -1 \\[1ex]  -1 & -1 & -1 & -1 \\[1ex]  -1 & -1 & -1 & -1 \\[-0.25ex] &&& \end{smallmatrix} \right]
\eeqa \end{subequations}

Note that $\bar{R}_{S,16}$ is distinguished from the other operations not only by the fact that it is not orientation-preserving, but also by the fact that it is nonlocal and hence does not preserve the tensor product structure in the space of (real or Hermitian) density matrices.
It is easily seen that the involutory mapping induced by any tensor product of sign matrices as above must preserve orientation, but there are many orientation-preserving mappings that are not tensor products, including the pair given below:
\beq
\left[ \begin{smallmatrix} &&& \\[0.5ex]  +1 & +1 & +1 & +1 \\[1ex]  +1 & -1 & -1 & +1 \\[1ex]  +1 & -1 & -1 & +1 \\[1ex]  +1 & +1 & +1 & +1 \\[-0.25ex] &&& \end{smallmatrix} \right]
~\longleftrightarrow~
\left[ \begin{smallmatrix} &&& \\[0.5ex]  +1 & +1 & +1 & -1 \\[1ex]  +1 & +1 & -1 & +1 \\[1ex]  +1 & -1 & +1 & +1 \\[1ex]  -1 & +1 & +1 & +1 \\[-0.25ex] &&& \end{smallmatrix} \right].
\eeq
As indicated by the double arrow, these two are related by the Choi map, i.e.~taking the Hadamard product of one with the real density matrix is the same as taking the Hadamard product of the other with the Stokes tensor (cf.~Eq.~(\ref{eq:tensor2matrix})).
Tests with randomly generated pure states quickly show that neither of these maps is positive, let alone completely positive.

Similarly, one can easily construct many other discrete reflection symmetries which are neither locally nor unitarily equivalent to the total reflection, simply by composing the latter with any other nonlocal and nonunitary rotation symmetry.
One interesting example is obtained by composing the local reflections $\bar{R}_S \otimes \bar{R}_S$ with the total reflection $\bar{R}_{S,16}$ on two qubits, obtaining
\beq
\big( \bar{R}_S \otimes \bar{R}_S \big) \bar{R}_{S,16} ~\leftrightarrow~ C ~\equiv~ \left[ \begin{smallmatrix} &&& \\[0.5ex]  +1 & +1 & +1 & -1 \\[1ex]  +1 & +1 & -1 & -1 \\[1ex]  +1 & -1 & +1 & -1 \\[1ex]  -1 & -1 & -1 & -1 \\[-0.25ex] &&& \end{smallmatrix} \right].
\eeq
The Hadamard product with $C$ changes the sign of the bilinear (two-body) part of the Stokes tensor. It is, of course, a non-positive map which takes the Hermitian density matrix of any pure state to one with eigenvalues $[1,1,1,-1]/2$.
This map may also be written quite simply as an operator sum, as follows:

\begin{multline}
\sigma^{-1} ( C \odot \sigma (\rho) ) ~=\cdots \\
\sum_{k=1}^{3} \big(  \Lambda^{k0} \rho\, \Lambda^{k0} + \Lambda^{0k} \rho\, \Lambda^{0k} \big) \,-\, \tfrac12\, \openone_4 ~.
\end{multline}

Finally, we show how a separability test based on the so-called computable cross-norm (CCN), denoted in what follows by ``$\xi$'', can be performed directly using the Stokes tensor.
The CCN is a lower bound on the cross-norm entanglement measure in a bipartite system, denoted by ``$\Xi$'', which satisfies $\Xi(\rho) = 1$ if $\rho$ is separable and $\Xi(\rho) > 1$ if it is not \cite{Rudolph}.
Consequently, $\xi(\rho) > 1$ implies $\rho$ is nonseparable, though not vice-versa; this condition is neither weaker nor stronger than the PPT criterion, but inequivalent to it.
The CCN $\xi$ is not itself an entanglement measure, since it may increase under the partial trace operation, but it has the advantage that it is readily computed as the sum of the singular values (aka trace class norm) of the reshuffled density matrix $\mathrm{Choi}(\rho)$.
For two qubits it can also be computed directly from the Stokes tensor, as shown by the following:

\begin{proposition}
For two qubits, the singular values of the Stokes tensor $\varrho^{k\ell}$, regarded as a matrix as in Eq.~\eqref{eq:tensor2matrix}, are twice those of the reshuffled density matrix $\mathrm{Choi}(\rho)$.
\end{proposition}
\proof
The reshuffling operation is defined to satisfy $\mathrm{Choi}(\rho_1^T \otimes \rho_2) = \ket{\rho_2}\bra{\rho_1^T}$, where $\ket{\rho_2}$ denotes the result of applying the reshaping operator to $\rho_2$, and $\bra{\rho_1^T} = \ket{\rho_1}^T$.
The one nonzero singular value of this matrix is simply the product of the Hilbert-Schmidt norms of its factors $\| \rho_1 \| \| \rho_2 \|$.
Recall now that $\rho$ is factorizable if and only if the corresponding real density matrix $\sigma(\rho)$ is and that the linear mapping $\sigma/2^{n/2}$ preserves the Hilbert-Schmidt norm (where $n$ is the number of qubits).
Hence $\ket{\sigma(\rho_2)}\bra{\sigma(\rho_1^T)}/2$ is the singular value decomposition of the corresponding reshuffled real density matrix $\mathrm{Choi}(\sigma(\rho_1^T \otimes \rho_2))/2$, and its nonzero singular value is $\| \sigma(\rho_1) \| \| \sigma(\rho_2) \| / 2 = \| \rho_1 \| \| \rho_2 \|$.
Together with the fact that for two qubits the Stokes tensor and the real density matrix are related by the $\mathrm{Choi}$ map, this establishes the result for factorizable states.

To prove the general case, we recall that the reshaping map $\mathrm{Choi}$ is self-inverse.
Thus the singular value decomposition of a general matrix $\mathrm{Choi}(\rho) = \sum_k \xi_k r_k s_k^T$ provides a canonical decomposition of $\rho$ into a sum of tensor products $\sum_k p_k\, \rho_{1k}^T \otimes \rho_{2k}$, where $p_k = \xi_k\, r_k^T \ket{\openone_2}\, s_k^T \ket{\openone_2}$.
Although $p_k$ may be negative and the factors $\rho_{1k}^T$, $\rho_{2k}$ of each term in this sum are not necessarily states (i.e.~nonnegative definite), we are free to apply the composition $\mathrm{Choi} \circ \sigma$ to each term $\rho_{1k}^T \otimes \rho_{2k}$ thereby obtained.
Then noting that $\sigma$ also preserves orthogonality and invoking the uniqueness of singular value decompositions completes the proof. \qed

The claim that $\rho$ is separable implies $\xi(\rho) \le 1$ can now be established directly, since $\xi( \rho_1 \otimes \rho_2 ) = \| \rho_1 \| \| \rho_2 \| \le 1$ and $\xi$ satisfies the triangle inequality just like any norm,  so that for any $p_k \ge 0$ summing to unity we have $\xi( \sum_k p_k\, \rho_{1k} \otimes \rho_{2k} ) \le \sum_k p_k = 1$.
The singular value decomposition of these matrices can be regarded as an extension of the Schmidt composition for pure states to mixed states.
Indeed it can be shown that for pure states $\mathrm{Choi}(\rho)$ has a degenerate pair of singular values which are equal to twice the product of the corresponding Schmidt coefficients.

\section{Reflections on three or\hspace{3em} more qubits}

The situation is similar for three (or more) qubits, since the adjoint action (conjugation) still corresponds to a real ``one-sided'' rotation of the Stokes tensor, and the rotation group in all dimensions splits into orientation preserving \& changing connected components.
The main difference is that the number of inequivalent kinds of rotations and reflections goes us rapidly with the number of qubits.
Indeed, there are $2^{3n}$ local symmetries (i.e.~sign changes in the Bloch vector components),
and dividing this into the total number of trace-preserving discrete symmetries gives
\beq
2^{4^n-1} \big/\, 2^{3n} ~=~ 2^{4^n-3n-1}
\eeq
locally inequivalent symmetry operations.

It is possible, however, to identify some particularly significant involutions. In the case of three qubits $ \rho = \varrho^{jkl} \Lambda_{jkl} $, the following are some of the new possibilities:
\begin{description}
\item[(ia)] the two-qubit partial transposition $ \bar{R}_{T,16}\ox \openone_4 $ (and the two others obtained by qubit permutation);
\item[(ib)] the total transposition $ \bar{R}_{T, 64} = \mathrm{diag} ( 1, R_{T,63} ) $ (where $ R_{T, 63} \in SO(63)$ is diagonal with 28 $-1$'s and 35 $+1$'s in it), which changes the sign of just those elements $\varrho^{jkl}$ with an odd number of indices equal ``$2$''; 
\item[(iia)] the two-qubit ``reflection''  $\bar{R}_{S,16}\ox \openone_4$ (and the two others obtained by qubit permutation) -- which is however an orientation-preserving rotation on three qubits; 
\item[(iib)] the total (three-qubit) reflection $\bar{R}_{S,64} = \mathrm{diag} \left(  1, \,-\openone_{63} \right)$.
\end{description}

The effect of $\bar{R}_{S,64}$ on $\varrho^{jkl}$ is to change the sign of the entire tensor except for its constant component $\varrho^{000} = 1/(2 \sqrt{2})$, showing that it may be expressed as:
\beq
\bar{R}_{S,64} (\rho) ~=~ 2 \varrho^{000} \Lambda_{000} -\rho ~=~ \tfrac14 \openone_8 - \rho .
\label{eq:3qubit-refl}
\eeq
This is again a nonlocal operation which admits no factorization into independent one-qubit operations.
Similarly, the action of $\bar{R}_{S,16}\ox \openone_4$ on $\varrho^{jk\ell}$ is to change the sign of the entire tensor except for the Bloch vector of the 1-qubit reduced density $ \varrho^{00\ell} $.
Items \textbf{(ia)} and \textbf{(ib)} above are fundamentally different from \textbf{(iia)} and \textbf{(iib)}.
The first two produce a Hermitian matrix with negative eigenvalues whenever the density has bipartite entanglement through the cut, whereas the latter two instead may map even separable densities to Hermitian matrices with negative eigenvalues.
This can be seen looking at the components of the UPB state.
If $ \bar{R}_{S,64} $ is applied to the (separable) density $ \rho_{\rm UPB-sep} = \frac{1}{4} \sum_{j=1}^4 \ket{\psi_j} \bra{\psi_j } $ with $ \ket{\psi_j} = \ket{01+}, \ket{1+0},\ket{+01}, \ket{---} $ and $\ket{\pm} = \frac{1}{\sqrt{2}} \left( \ket{0} \pm \ket{1} \right) $, one gets the bound entangled state $ \rho_{\rm UPB} $ used in \cite{Bennett1}.
So in this case a separable state is reflected into an entangled state.
However, no one of the 4 components $ \ket{\psi_j}\bra{\psi_j}$ taken alone (each is obviously separable) is a density when reflected.
Obviously (ib) only reverses the time arrow on any 3-qubit density.

In similar fashion, for any number $n>1$ of qubits one can define an \emph{$m$-qubit} ($1<m\leqslant n$) \emph{nonlocal} ``reflection'' $\bar{R}_{S,4^m} \ox \openone_{4^{n-m}}$, which is only a true (i.e.~orientation-changing) reflection when the reflection is total ($m = n$).
Assuming the reflection acts on the first $m$ qubits of an $n$-qubit density operator $\rho$, this may be written as
\beq
\hspace{-0.25em} \bar{R}_{S,4^m} \ox \openone_{4^{n-m}} (\rho ) \,=\, 2 \varrho^{0\ldots 0 j_{m+1} \ldots j_n } \Lambda_{0\ldots 0 j_{m+1} \ldots j_n } - \rho \,. \hspace{-0.25em}
\label{eq:part-refl-n}
\eeq
These operations leave the norm of the $n$-qubit tensor $ \varrho^{j_1 \ldots j_n}$ (i.e.~$\tr{\rho^2}$) invariant, but need not preserve the spectrum nor even leave it nonnegative, as we saw above.
Hence it is a ``generically'' ill-defined operation on the set of density operators of composite systems ${\cal D}_n$.

These observations are summarized in the following:
\begin{proposition}
In ${\cal D}_n$, the linear map $\bar{R}_{S,4^n}$ $(1 <  n)$:
\begin{description}
\item[(i)] preserves the trace and Hermiticity;
\item[(ii)] preserves the Hilbert-Schmidt inner product;
\item[(iii)] is neither unitary nor antiunitary; 
\item[(iv)] is not $ {\cal D}_n$-invariant.
\end{description}
\end{proposition}
Properties \textbf{(i)} and \textbf{(ii)} together say that $\bar{R}_{S,4^n}$ is neither a contraction nor a dilation map, whereas \textbf{(iv)} affirms that $ \bar{R}_{S,4^n} $ is not a positive map.

Nevertheless, it is possible to specify a simple spectral condition on the density matrix that guarantees that its total reflection is still nonnegative definite.
\begin{theorem}
\label{prop:suff-cond-be-refl}
Given $\rho = \varrho^{0\ldots 0} \Lambda_{0\ldots 0} + \chi  \in {\cal D}_n$ (where $\chi$ is the associated homogeneous tensor), a sufficient condition for $\tilde{\rho} = \bar{R}_{S,4^n} ( \rho ) = \varrho^{0\ldots 0} \Lambda_{0\ldots 0} - \chi \in {\cal D}_n$ is that the set of eigenvalues satisfies $\mathrm{eig} ( \rho ) \subset [ 0, \, 2^{1- n} ]$.
\end{theorem}
\proof
The proof is based on the well-known fact \cite{Horn1} that adding a multiple of the identity $c \openone_m$ onto an $m\times m$ Hermitian matrix $A$ shifts its eigenvalues by $c$, i.e.~$\mathrm{eig}( A + c \openone_m ) = \mathrm{eig}( A ) + c$.
Since the eigenvalues of the random state's density matrix $\mathrm{eig}\left(  \varrho^{0\ldots 0} \Lambda_{0\ldots 0} \right) = \{ 2^{-n} \}$ (with multiplicity $2^n$), we see that $\mathrm{eig} ( \rho ) \in [ 0, \, 2^{1-n} ]$ implies both $\mathrm{eig} ( \chi ) \in [-2^{-n},  \,2^{-n} ]$ and $\mathrm{eig} (- \chi ) \in [-2^{-n},  \,2^{-n} ]$, so that $ \mathrm{eig} ( \tilde{\rho} ) \in [ 0, \, 2^{1-n} ]$, as well.
\qed
\noindent Hence the linear map $ \bar{R}_{S,4^n} $ is well-defined (positive) in the subset of densities with eigenvalues in the interval $ [0,\, 2^{1-n} ] $.

\begin{corollary} 
\label{coroll:nec-cond-be-refl2}
A necessary but not sufficient condition for Theorem~\ref{prop:suff-cond-be-refl} to hold is that $ \tr{ \rho^2 } \leqslant2^{1-n} $.
\end{corollary}
\proof
If $ \mu_{\chi_1}$, $ \ldots $, $  \mu_{\chi_{2^n}}$ are the eigenvalues of $\chi$, when Theorem~\ref{prop:suff-cond-be-refl} holds it must be $ r^2= \mu_{\chi_1}^2 + \ldots + \mu_{\chi_{2^n}}^2 \leqslant \frac{1}{2^n} $. 
Hence $ r\leqslant \frac{1}{2^{n/2}} $ and $ \tr{\rho^2 } = (\varrho^{0\ldots 0})^2 + r^2 \leqslant   \frac{1}{2^{n-1}} $.
\qed
\begin{corollary} 
\label{coroll:nec-cond-be-refl}
A necessary but not sufficient condition for $ \tilde{\rho} $ to be a density is that $ \mathrm{rank} ( \rho ) \geqslant 2^{n-1} $.
\end{corollary}
In fact, only when $ \rho $ is a mixture of at least $ 2^{n-1} $ pure states one can achieve $ \mathrm{eig}( \rho) \in [ 0, \, \frac{1}{2^{n-1} } ] $.

\begin{figure}[ht]
\begin{center}
 \includegraphics[width=8.5cm]{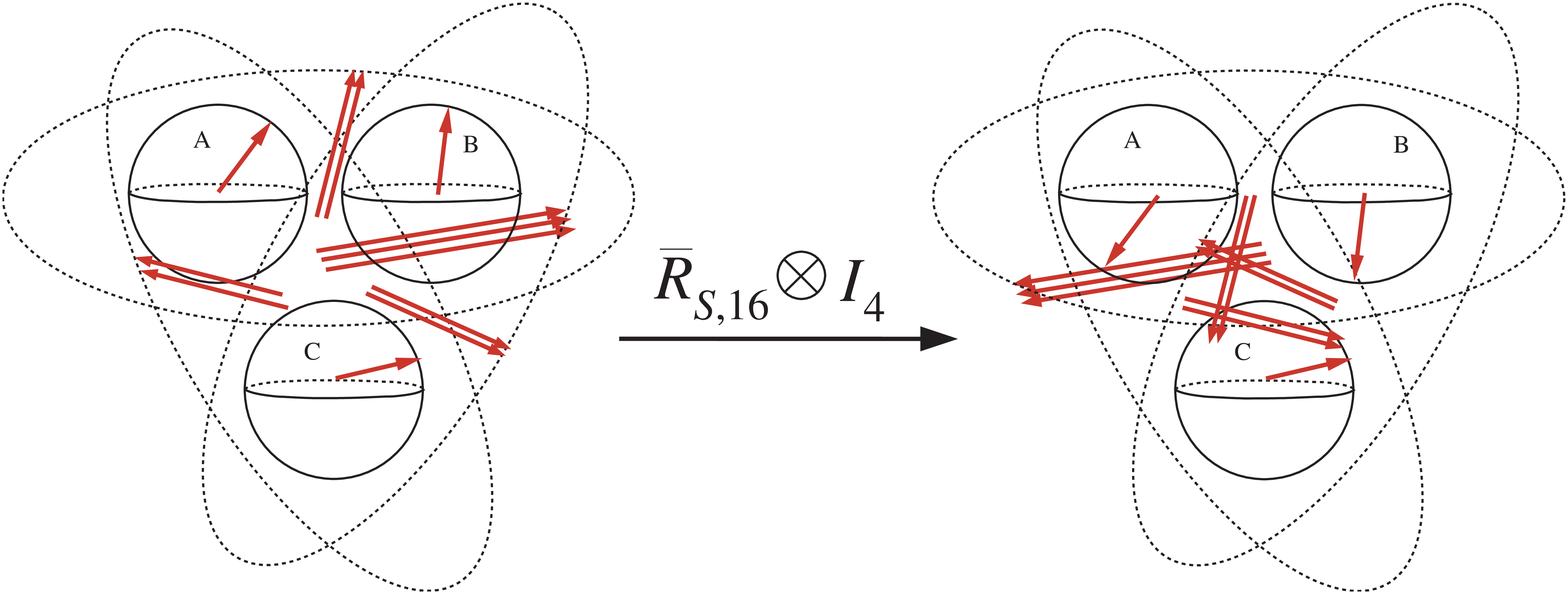}
 \caption{Schematic illustration of the action of a two-qubit reflection $\bar{R}_{S,16} \ox \openone_4 (\rho)$ on three qubits. One-body ($\varrho^{j00}$, $ \varrho^{0k0}$ and $ \varrho^{00\ell}$), two-body ($\varrho^{jk0}$, $\varrho^{j0\ell}$ and $\varrho^{0k\ell}$) and three-body ($\varrho^{jk\ell}$) correlation terms are indicated by single, double and triple arrows. All the signs are changed except for those of $\varrho^{00\ell}$.}
\label{fig:Stokes-ref4}
\end{center}
\end{figure}

On a 3-qubit density, the action of $\bar{R}_{S,16}\ox \openone_4$ is depicted in Fig.~\ref{fig:Stokes-ref4}.
Essentially the entire Stokes tensor changes sign, except for the reduced density $\mathrm{tr}_{AB} (\rho)$.
Its action closely resembles the reduction criterion of \cite{Cerf1,Horodecki8}.
That criterion also makes implicit use of nonlocal reflections, but it is formulated based on a positive map, hence it is well-posed on all of ${\cal D}_n$.
For a 3-qubit density it affirms that a necessary condition for separability is $\openone_2 \ox \openone_2 \ox  \mathrm{tr}_{AB} (\rho) - \rho \geqslant 0$ as well as $\openone_2 \ox \mathrm{tr}_{A}(\rho) \geqslant 0$ (and likewise for the other indexes).
Since $ \tr{\openone_2 \ox \openone_2 \ox  \mathrm{tr}_{AB} (\rho) - \rho} = 3$, one difference between our partial reflection and the reduction criterion is that the latter is not a trace preserving map.
Thus it is not a symmetry in the sense used in the paper.
Nonetheless, the reduction criterion utilizes a positive map which can be used to test separability.
In our case  $ \bar{R}_{S,16} \ox \openone_4 $ is not a positive map even when restricted to separable states with eigenvalues in $ [0,\, 1/4]$, in which the 3-qubit total reflection $ \bar{R}_{S,64} $ is always well-posed.

We can, however, convert our 2-qubit total reflection $\bar{R}_{S,16}$ into a ``relaxed''  total reflection, namely
\beq
\bar{R}_{S,16}^{rel}(\rho) ~=~ \tfrac13 \big( \openone_4 - \rho \big) ,
\eeq
which is the same as $\bar{R}_{S,16}$ applied to the ``remixed'' density matrix $(\openone_4 / 2 +  \rho) / 3$.
Since the remixed density matrix now has eigenvalues in $[0,1/2]$, the relaxed reflection is a positive map by Theorem \ref{prop:suff-cond-be-refl}.
It is also easily shown that $\bar{R}_{S,16}^{rel}$ is not completely positive, and hence provides a necessary condition for the separability of an arbitrary $2^n\times2^n$, $ n>2$, density matrix.
It should be possible to relax all the reflections described in this paper to positive maps by a similar strategy.

Concerning a total reflection, all pairs $ \rho $ and $ \tilde{\rho} = \bar{R}_{S,4^n} ( \rho )$ satisfying Theorem~\ref{prop:suff-cond-be-refl} are complementary in the sense that their mixture is the random state:
\beq
\frac{1}{2} \left( \rho + \tilde{\rho} \right) = \frac{1}{2^n} \openone_{2^n} .
\label{eq:mixrho-tilrho}
\eeq
Eq. \eqref{eq:mixrho-tilrho} implies that $ \bar{R}_{S,4^n} $ corresponds to a multiparty NOT operation.
In fact, also in the single qubit case, the NOT operation corresponds to a change of sign to the homogeneous part (i.e., the Bloch vector) but it is not modifying the sign of the trace part and hence a qubit and its reflection obey to \eqref{eq:mixrho-tilrho}.
Such operation is used for example to map a density operator belonging to a subset of the Hilbert space $ {\cal D}_n $ to its complement in $ {\cal D}_n $, for example in the UPB construction mentioned above \cite{Horodecki2,Bennett1}.

\section{Concluding Remarks}

Reflections are a natural discrete class of transformations relative to the Stokes tensor / real density matrix parametrization. Their meaning and relation to LOCC is interesting and calls for natural generalizations to nonlocal operations in the way explained above.
The nonlocal reflections, in fact, originate from the nonconnectedness of the group of rotations acting on the Stokes tensor parametrization.
In terms of density matrices, this interpretation is not as sharp. 
As a matter of fact, operations reducible to reflections appear in the PPT test and in the various measures of entanglement relying upon ``spin-flip'' operations (like concurrence, negativity and tangle) for what concerns (multiple) 1-qubit reflections. Also nonlocal reflections are used: for example a total reflection corresponds to what is normally referred to as ``taking the complement of a density'', used for example in the construction of Unextendible Product Basis states \cite{Bennett1}. 
Likewise, the reduction critetion makes use of a positive map closely related to our nonlocal reflections.

For the purposes of further understanding the structure of composite quantum systems, we find it useful to have a unifying perspective on these nonunitary yet symmetric (in the sense of Wigner Theorem) transformations.

It is worth pointing out that reflections can be defined in the same terms also for SLOOC (stochastic LOCC) \cite{Verstraete1,Jaeger1,Teodorescu1}.
For the Stokes tensor, in fact, this class of operations relaxes the group of admissible local transformations from affine rotations to proper orthochronous local Lorentz transformations $ SO(1,3) $. 
The reflected action in $ O^-(3)  $ then corresponds to choosing the other connected component of $ O(1,3)$ with the same time direction as $ SO(1,3) $ (i.e., with positive ``time-like'' metric element).
Also nonlocal reflections fit in with the group structure of nonlocal filtering operations. For example, total reflections belong to $ O(1,4^n-1)\setminus SO(1,4^n-1 )$.
Note further that the idea of restricting the set of density operators in order to have a larger set of symmetries is ``dual'' to the idea of using group actions that are contractions \cite{Horodecki5,Kossakowski1}.

The idea of using reflections does not extend in a straightforward manner to qutrids (nor to higher dimension quantum systems), as in this case the admissible parameters live on a rather complicated subset of the seven dimensional sphere \cite{Byrd2,Kimura1} for which the rotation representing transposition is always admissible but spatial inversion may not be.
However, the various UPB constructions on $ 3\times 3 $ systems of Ref.\ \cite{Bennett1} correspond to well-defined reflections.

Finally notice that there are many isometries of the Stokes tensor that do not correspond to reflections relative to any basis; those that do are of course involutions, and it would be interesting to show that any Stokes tensor isometry which is an involution (and hence is described by a symmetric orthogonal matrix) is a reflection relative to some basis.

\bibliographystyle{apsrev}
\small


\begin{thebibliography}{28}
\expandafter\ifx\csname natexlab\endcsname\relax\def\natexlab#1{#1}\fi
\expandafter\ifx\csname bibnamefont\endcsname\relax
  \def\bibnamefont#1{#1}\fi
\expandafter\ifx\csname bibfnamefont\endcsname\relax
  \def\bibfnamefont#1{#1}\fi
\expandafter\ifx\csname citenamefont\endcsname\relax
  \def\citenamefont#1{#1}\fi
\expandafter\ifx\csname url\endcsname\relax
  \def\url#1{\texttt{#1}}\fi
\expandafter\ifx\csname urlprefix\endcsname\relax\def\urlprefix{URL }\fi
\providecommand{\bibinfo}[2]{#2}
\providecommand{\eprint}[2][]{\url{#2}}

\bibitem[{\citenamefont{Sakurai}(1994)}]{Sakurai1}
\bibinfo{author}{\bibfnamefont{J.~J.} \bibnamefont{Sakurai}},
  \emph{\bibinfo{title}{Modern Quantum Mechanics}}
  (\bibinfo{publisher}{Addison-Wesley Longman}, \bibinfo{year}{1994}).

\bibitem[{\citenamefont{Buzek et~al.}(1999)\citenamefont{Buzek, Hillery, and
  Werner}}]{Buzek1}
\bibinfo{author}{\bibfnamefont{V.}~\bibnamefont{Buzek}},
  \bibinfo{author}{\bibfnamefont{H.}~\bibnamefont{Hillery}}, \bibnamefont{and}
  \bibinfo{author}{\bibfnamefont{R.~F.} \bibnamefont{Werner}},
  \bibinfo{journal}{Phys. Rev. A} \textbf{\bibinfo{volume}{60}},
  \bibinfo{pages}{R2626} (\bibinfo{year}{1999}).

\bibitem[{\citenamefont{Gisin and Popescu}(1999)}]{Gisin1}
\bibinfo{author}{\bibfnamefont{N.}~\bibnamefont{Gisin}} \bibnamefont{and}
  \bibinfo{author}{\bibfnamefont{S.}~\bibnamefont{Popescu}},
  \bibinfo{journal}{Phys. Rev. Lett.} \textbf{\bibinfo{volume}{83}},
  \bibinfo{pages}{432} (\bibinfo{year}{1999}).

\bibitem[{\citenamefont{Rungta et~al.}(2001)\citenamefont{Rungta, Buzek, Caves,
  Hillery, and Milburn}}]{Rungta1}
\bibinfo{author}{\bibfnamefont{P.}~\bibnamefont{Rungta}},
  \bibinfo{author}{\bibfnamefont{V.}~\bibnamefont{Buzek}},
  \bibinfo{author}{\bibfnamefont{C.~M.} \bibnamefont{Caves}},
  \bibinfo{author}{\bibfnamefont{M.}~\bibnamefont{Hillery}}, \bibnamefont{and}
  \bibinfo{author}{\bibfnamefont{G.~J.} \bibnamefont{Milburn}},
  \bibinfo{journal}{Phys. Rev. A} \textbf{\bibinfo{volume}{64}},
  \bibinfo{pages}{042315} (\bibinfo{year}{2001}).

\bibitem[{\citenamefont{Doran \& Lasenby}(2004)}]{DoranLasenby}
\bibinfo{author}{\bibfnamefont{C.~J.~L.} \bibnamefont{Doran}}, \bibnamefont{and}
\bibinfo{author}{\bibfnamefont{A.~N.} \bibnamefont{Lasenby}},
  \emph{\bibinfo{title}{Geometric Algebra for Physicists}}
  (\bibinfo{publisher}{Cambridge University Press}, \bibinfo{year}{2004}).

\bibitem[{\citenamefont{Peres}(1996)}]{Peres1}
\bibinfo{author}{\bibfnamefont{A.}~\bibnamefont{Peres}},
  \bibinfo{journal}{Phys. Rev. Lett.} \textbf{\bibinfo{volume}{77}},
  \bibinfo{pages}{1413} (\bibinfo{year}{1996}).

\bibitem[{\citenamefont{Altafini}(2004{\natexlab{a}})}]{Cla-qu-ent1}
\bibinfo{author}{\bibfnamefont{C.}~\bibnamefont{Altafini}},
  \bibinfo{journal}{Phys. Rev. A} \textbf{\bibinfo{volume}{69}},
  \bibinfo{pages}{012311} (\bibinfo{year}{2004}{\natexlab{a}}).

\bibitem[{\citenamefont{Altafini}(2004{\natexlab{b}})}]{Cla-spin-tens1}
\bibinfo{author}{\bibfnamefont{C.}~\bibnamefont{Altafini}},
  \bibinfo{journal}{Physical Review A} \textbf{\bibinfo{volume}{70}},
  \bibinfo{pages}{032331} (\bibinfo{year}{2004}{\natexlab{b}}).

\bibitem[{\citenamefont{Jaeger et~al.}(2003{\natexlab{a}})\citenamefont{Jaeger,
  Teodorescu-Frumosu, Sergienko, Saleh, and Teich}}]{Jaeger1}
\bibinfo{author}{\bibfnamefont{G.}~\bibnamefont{Jaeger}},
  \bibinfo{author}{\bibfnamefont{M.}~\bibnamefont{Teodorescu-Frumosu}},
  \bibinfo{author}{\bibfnamefont{A.}~\bibnamefont{Sergienko}},
  \bibinfo{author}{\bibfnamefont{B.~E.~A.} \bibnamefont{Saleh}},
  \bibnamefont{and} \bibinfo{author}{\bibfnamefont{M.~C.} \bibnamefont{Teich}},
  \bibinfo{journal}{Phys. Rev. A} \textbf{\bibinfo{volume}{67}},
  \bibinfo{pages}{032307} (\bibinfo{year}{2003}{\natexlab{a}}).

\bibitem[{\citenamefont{Havel}(2003a)}]{Havel3}
\bibinfo{author}{\bibfnamefont{T.~F.} \bibnamefont{Havel}},
  \bibinfo{journal}{Q. Inf. Proc.} \textbf{\bibinfo{volume}{1}},
  \bibinfo{pages}{511} (\bibinfo{year}{2003}).

\bibitem[{\citenamefont{Alicki and Lendi}(1987)}]{Alicki1}
\bibinfo{author}{\bibfnamefont{R.}~\bibnamefont{Alicki}} \bibnamefont{and}
  \bibinfo{author}{\bibfnamefont{K.}~\bibnamefont{Lendi}},
  \emph{\bibinfo{title}{Quantum Dynamical Semigroups and Applications}}, vol.
  \bibinfo{volume}{286} of \emph{\bibinfo{series}{Lecture Notes in Physics}}
  (\bibinfo{publisher}{Springer-Verlag}, \bibinfo{year}{1987}).

\bibitem[{\citenamefont{Mahler and Weberruss}(1998)}]{MahlerWeberruss}
\bibinfo{author}{\bibfnamefont{G.}~\bibnamefont{Mahler}} \bibnamefont{and}
  \bibinfo{author}{\bibfnamefont{V.~A.}~\bibnamefont{Weberruss}},
  \emph{\bibinfo{title}{Quantum Networks (2nd edition)}},
  (\bibinfo{publisher}{Springer-Verlag}, \bibinfo{year}{1998}).

\bibitem[{\citenamefont{Bennett et~al.}(1999)\citenamefont{Bennett, DiVincenzo,
  Mor, Shor, Smolin, and Terhal}}]{Bennett1}
\bibinfo{author}{\bibfnamefont{C.~H.} \bibnamefont{Bennett}},
  \bibinfo{author}{\bibfnamefont{D.~P.} \bibnamefont{DiVincenzo}},
  \bibinfo{author}{\bibfnamefont{T.}~\bibnamefont{Mor}},
  \bibinfo{author}{\bibfnamefont{P.~W.} \bibnamefont{Shor}},
  \bibinfo{author}{\bibfnamefont{J.~A.} \bibnamefont{Smolin}},
  \bibnamefont{and} \bibinfo{author}{\bibfnamefont{B.~M.}
  \bibnamefont{Terhal}}, \bibinfo{journal}{Phys. Rev. Lett.}
  \textbf{\bibinfo{volume}{82}}, \bibinfo{pages}{5385} (\bibinfo{year}{1999}).

\bibitem[{\citenamefont{Cerf et~al.}(1999)\citenamefont{Cerf, Adami, and
  Gingrich}}]{Cerf1}
\bibinfo{author}{\bibfnamefont{N.~J.} \bibnamefont{Cerf}},
  \bibinfo{author}{\bibfnamefont{C.}~\bibnamefont{Adami}}, \bibnamefont{and}
  \bibinfo{author}{\bibfnamefont{R.~M.} \bibnamefont{Gingrich}},
  \bibinfo{journal}{Phys. Rev. A} \textbf{\bibinfo{volume}{60}},
  \bibinfo{pages}{898} (\bibinfo{year}{1999}).

\bibitem[{\citenamefont{Horodecki and Horodecki}(1999)}]{Horodecki8}
\bibinfo{author}{\bibfnamefont{M.}~\bibnamefont{Horodecki}} \bibnamefont{and}
  \bibinfo{author}{\bibfnamefont{P.}~\bibnamefont{Horodecki}},
  \bibinfo{journal}{Phys. Rev. A} \textbf{\bibinfo{volume}{59}},
  \bibinfo{pages}{4206} (\bibinfo{year}{1999}).

\bibitem[{\citenamefont{Havel}(2003b)}]{Havel2}
\bibinfo{author}{\bibfnamefont{T.~F.} \bibnamefont{Havel}},
  \bibinfo{journal}{J.~Math.~Phys.} \textbf{\bibinfo{volume}{44}},
  \bibinfo{pages}{534} (\bibinfo{year}{2003}).

\bibitem[{\citenamefont{Zyczkowski and Bengtsson}(2004)}]{ZyczkowskiBengtsson}
\bibinfo{author}{\bibfnamefont{K.} \bibnamefont{Zyczkowski}},  \bibnamefont{and}
\bibinfo{author}{\bibfnamefont{I.} \bibnamefont{Bengtsson}},
  \bibinfo{journal}{Open Sys.~\& Inform.~Dyn.} \textbf{\bibinfo{volume}{11}},
  \bibinfo{pages}{3} (\bibinfo{year}{2004}).

\bibitem[{\citenamefont{Busch and Lahti}(1997)}]{Busch1}
\bibinfo{author}{\bibfnamefont{P.}~\bibnamefont{Busch}} \bibnamefont{and}
  \bibinfo{author}{\bibfnamefont{P.~J.} \bibnamefont{Lahti}},
  \bibinfo{journal}{Found. Phys. Lett.} \textbf{\bibinfo{volume}{10}},
  \bibinfo{pages}{113} (\bibinfo{year}{1997}).

\bibitem[{\citenamefont{Lewenstein et~al.}(2000)\citenamefont{Lewenstein,
  Bruss, Cirac, Kraus, Kus, Samsonowicz, Sanpera, and Tarrach}}]{Lewenstein1}
\bibinfo{author}{\bibfnamefont{M.}~\bibnamefont{Lewenstein}},
  \bibinfo{author}{\bibfnamefont{D.}~\bibnamefont{Bruss}},
  \bibinfo{author}{\bibfnamefont{J.~I.} \bibnamefont{Cirac}},
  \bibinfo{author}{\bibfnamefont{B.}~\bibnamefont{Kraus}},
  \bibinfo{author}{\bibfnamefont{M.}~\bibnamefont{Kus}},
  \bibinfo{author}{\bibfnamefont{J.}~\bibnamefont{Samsonowicz}},
  \bibinfo{author}{\bibfnamefont{A.}~\bibnamefont{Sanpera}}, \bibnamefont{and}
  \bibinfo{author}{\bibfnamefont{R.}~\bibnamefont{Tarrach}},
  \bibinfo{journal}{Journal of Modern Optics} \textbf{\bibinfo{volume}{47}},
  \bibinfo{pages}{2481} (\bibinfo{year}{2000}).

\bibitem[{\citenamefont{Jaeger et~al.}(2003{\natexlab{b}})\citenamefont{Jaeger,
  Sergienko, Saleh, and Teich}}]{Jaeger2}
\bibinfo{author}{\bibfnamefont{G.}~\bibnamefont{Jaeger}},
  \bibinfo{author}{\bibfnamefont{A.~V.} \bibnamefont{Sergienko}},
  \bibinfo{author}{\bibfnamefont{B.~E.~A.} \bibnamefont{Saleh}},
  \bibnamefont{and} \bibinfo{author}{\bibfnamefont{M.~C.} \bibnamefont{Teich}},
  \bibinfo{journal}{Phys. Rev. A} \textbf{\bibinfo{volume}{68}},
  \bibinfo{pages}{022318} (\bibinfo{year}{2003}{\natexlab{b}}).

\bibitem[{\citenamefont{Ruskai et~al.}(2002)\citenamefont{Ruskai, Szarek, and
  Werner}}]{Ruskai1}
\bibinfo{author}{\bibfnamefont{M.~B.} \bibnamefont{Ruskai}},
  \bibinfo{author}{\bibfnamefont{S.}~\bibnamefont{Szarek}}, \bibnamefont{and}
  \bibinfo{author}{\bibfnamefont{E.}~\bibnamefont{Werner}},
  \bibinfo{journal}{Lin. Alg. Appl.} \textbf{\bibinfo{volume}{347}},
  \bibinfo{pages}{159} (\bibinfo{year}{2002}).

\bibitem[{\citenamefont{Altafini}(2003)}]{Cla-contr-open1}
\bibinfo{author}{\bibfnamefont{C.}~\bibnamefont{Altafini}},
  \bibinfo{journal}{Journal of Mathematical Physics}
  \textbf{\bibinfo{volume}{44}}, \bibinfo{pages}{2357} (\bibinfo{year}{2003}).

\bibitem[{\citenamefont{Horodecki et~al.}(1996)\citenamefont{Horodecki,
  Horodecki, and Horodecki}}]{Horodecki1}
\bibinfo{author}{\bibfnamefont{M.}~\bibnamefont{Horodecki}},
  \bibinfo{author}{\bibfnamefont{P.}~\bibnamefont{Horodecki}},
  \bibnamefont{and}
  \bibinfo{author}{\bibfnamefont{R.}~\bibnamefont{Horodecki}},
  \bibinfo{journal}{Phys. Lett. A} \textbf{\bibinfo{volume}{223}},
  \bibinfo{pages}{1} (\bibinfo{year}{1996}).

\bibitem[{\citenamefont{Wootters}(2001)}]{Wootters2}
\bibinfo{author}{\bibfnamefont{W.~K.} \bibnamefont{Wootters}},
  \bibinfo{journal}{Quant. Inf. Comp.} \textbf{\bibinfo{volume}{1}},
  \bibinfo{pages}{27} (\bibinfo{year}{2001}).

\bibitem[{\citenamefont{Horn and Johnson}(1985)}]{Horn1}
\bibinfo{author}{\bibfnamefont{R.}~\bibnamefont{Horn}} \bibnamefont{and}
  \bibinfo{author}{\bibfnamefont{C.~R.} \bibnamefont{Johnson}},
  \emph{\bibinfo{title}{Matrix Analysis}} (\bibinfo{publisher}{Cambdridge
  University Press}, \bibinfo{year}{1985}).

\bibitem[{\citenamefont{Rudolph}(2003)}]{Rudolph}
\bibinfo{author}{\bibfnamefont{O. Rudolph}},
   \bibinfo{journal}{quant-ph/0212047v2}
   (\bibinfo{year}{2003}).

\bibitem[{\citenamefont{Chen and Wu}(2003)}]{Chen2}
\bibinfo{author}{\bibfnamefont{K.}~\bibnamefont{Chen}} \bibnamefont{and}
  \bibinfo{author}{\bibfnamefont{L.}~\bibnamefont{Wu}},
  \bibinfo{journal}{Quantum Information and Computation}
  \textbf{\bibinfo{volume}{3}}, \bibinfo{pages}{193} (\bibinfo{year}{2003}).

\bibitem[{\citenamefont{Horodecki}(1997)}]{Horodecki2}
\bibinfo{author}{\bibfnamefont{P.}~\bibnamefont{Horodecki}},
  \bibinfo{journal}{Phys. Lett. A} \textbf{\bibinfo{volume}{232}},
  \bibinfo{pages}{333} (\bibinfo{year}{1997}).

\bibitem[{\citenamefont{Verstraete et~al.}(2001)\citenamefont{Verstraete,
  Dehaene, and DeMoor}}]{Verstraete1}
\bibinfo{author}{\bibfnamefont{F.}~\bibnamefont{Verstraete}},
  \bibinfo{author}{\bibfnamefont{J.}~\bibnamefont{Dehaene}}, \bibnamefont{and}
  \bibinfo{author}{\bibfnamefont{B.}~\bibnamefont{DeMoor}},
  \bibinfo{journal}{Phys. Rev. A} \textbf{\bibinfo{volume}{64}},
  \bibinfo{pages}{010101} (\bibinfo{year}{2001}).

\bibitem[{\citenamefont{Teodorescu-Frumosu and Jaeger}(2003)}]{Teodorescu1}
\bibinfo{author}{\bibfnamefont{M.}~\bibnamefont{Teodorescu-Frumosu}}
  \bibnamefont{and} \bibinfo{author}{\bibfnamefont{G.}~\bibnamefont{Jaeger}},
  \bibinfo{journal}{Phys. Rev. A} \textbf{\bibinfo{volume}{67}},
  \bibinfo{pages}{052305} (\bibinfo{year}{2003}).

\bibitem[{\citenamefont{Byrd and Khaneja}(2003)}]{Byrd2}
\bibinfo{author}{\bibfnamefont{M.}~\bibnamefont{Byrd}} \bibnamefont{and}
  \bibinfo{author}{\bibfnamefont{N.}~\bibnamefont{Khaneja}},
  \bibinfo{journal}{Phys. Rev. A} \textbf{\bibinfo{volume}{68}},
  \bibinfo{pages}{062322} (\bibinfo{year}{2003}).

\bibitem[{\citenamefont{Kimura}(2003)}]{Kimura1}
\bibinfo{author}{\bibfnamefont{G.}~\bibnamefont{Kimura}},
  \bibinfo{journal}{Phys. Lett. A} \textbf{\bibinfo{volume}{314}},
  \bibinfo{pages}{339} (\bibinfo{year}{2003}).

\bibitem[{\citenamefont{Horodecki et~al.}(2002)\citenamefont{Horodecki,
  Horodecki, and Horodecki}}]{Horodecki5}
\bibinfo{author}{\bibfnamefont{M.}~\bibnamefont{Horodecki}},
  \bibinfo{author}{\bibfnamefont{P.}~\bibnamefont{Horodecki}},
  \bibnamefont{and}
  \bibinfo{author}{\bibfnamefont{R.}~\bibnamefont{Horodecki}},
  \bibinfo{journal}{quant-ph/0206008}  (\bibinfo{year}{2002}).

\bibitem[{\citenamefont{Kossakowski}(2003)}]{Kossakowski1}
\bibinfo{author}{\bibfnamefont{A.}~\bibnamefont{Kossakowski}},
  \bibinfo{journal}{Open Syst. Inform. Dyn.} \textbf{\bibinfo{volume}{10}},
  \bibinfo{pages}{213} (\bibinfo{year}{2003}).

\end{thebibliography}

\end{document}